# CONSTRUCTION SAFETY RISK MODELING AND SIMULATION

Antoine J.-P. Tixier[1], Matthew R. Hallowell[2], Balaji Rajagopalan[3]

[1]Postdoctoral researcher, Computer Science Laboratory, École Polytechnique, Palaiseau, France; antoine.tixier-1@colorado.edu
[2]Associate Professor and Beavers Faculty Fellow, Department of Civil, Environmental, and Architectural Engineering, CU Boulder, USA; matthew.hallowell@colorado.edu
[3]Professor and Chair, Department of Civil, Environmental, and Architectural Engineering, CU Boulder, USA; balajir@colorado.edu

## ABSTRACT

By building on a recently introduced genetic-inspired attribute-based conceptual framework for safety risk analysis, we propose a novel methodology to compute construction univariate and bivariate construction safety risk at a situational level. Our fully data-driven approach provides construction practitioners and academicians with an easy and automated way of extracting valuable empirical insights from databases of unstructured textual injury reports. By applying our methodology on an attribute and outcome dataset directly obtained from 814 injury reports, we show that the frequency-magnitude distribution of construction safety risk is very similar to that of natural phenomena such as precipitation or earthquakes. Motivated by this observation, and drawing on state-of-the-art techniques in hydroclimatology and insurance, we introduce univariate and bivariate nonparametric stochastic safety risk generators, based on Kernel Density Estimators and Copulas. These generators enable the user to produce large numbers of synthetic safety risk values faithfully to the original data, allowing safety-related decision-making under uncertainty to be grounded on extensive empirical evidence. Just like the accurate modeling and simulation of natural phenomena such as wind or streamflow is indispensable to successful structure dimensioning or water reservoir management, we posit that improving construction safety calls for the accurate modeling, simulation, and assessment of safety risk. The underlying assumption is that like natural phenomena, construction safety may benefit from being studied in an empirical and quantitative way rather than qualitatively which is the current industry standard. Finally, a side but interesting finding is that attributes related to high energy levels (e.g., machinery, hazardous substance) and to human error (e.g., improper security of tools) emerge as strong risk shapers on the dataset we used to illustrate our methodology.

## INTRODUCTION AND MOTIVATION

Despite the significant improvements in safety that have followed the inception of the Occupational Safety and Health Act of 1970, safety performance has reached a plateau in recent years and construction still accounts for a disproportionate accident rate. From 2013 to 2014, fatalities in construction even increased by 5% to reach 885, the highest count since 2008 (Bureau of Labor Statistics 2015). In addition to terrible human costs, construction injuries are also associated with huge direct and indirect economic impacts.

Partly due to their limited personal history with accidents, even the most experienced workers and safety managers may miss hazards and underestimate the risk of a given construction situation (Albert et al. 2014, Carter and Smith 2006). Designers face an even greater risk of failing to recognize hazards and misestimating risk (Albert et al. 2014, Almén and Larsson 2012). Therefore, a very large portion of construction work, upstream or downstream of ground-breaking, involves safety-related decision-making under uncertainty. Unfortunately, even more when uncertainty is involved, humans often recourse to personal opinion and intuition to apprehend their environment. This process is fraught with numerous biases and misconceptions inherent to human cognition (e.g., Kahneman and Tversky 1982) and compounds the likelihood of misdiagnosing the riskiness of a situation.

Therefore, it is of paramount importance to provide construction practitioners with tools to mitigate the adverse consequences of uncertainty on their safety-related decisions. In this study, we focus on leveraging situational data extracted from raw textual injury reports to guide and improve construction situation risk assessment. Our methodology facilitates the augmentation of construction personnel's experience and grounds risk assessment on potentially unlimited amounts of empirical and objective data. Put differently, our approach combats construction risk misdiagnosis on two fronts, by jointly addressing both the limited personal history and the judgment bias problems previously evoked.

We leveraged attribute data extracted by a highly accurate Natural Language Processing (NLP) system (Tixier et al. 2016a) from a database of 921 injury reports provided by a partner organization engaged in industrial construction projects worldwide.

Fundamental construction attributes are context-free universal descriptors of the work environment. They are observable prior to injury occurrence and relate to environmental conditions, construction means and methods, and human factors. To illustrate, one can extract four attributes from the following text: "worker is unloading a ladder from pickup truck with bad posture": ladder, manual handling, light vehicle, and improper body positioning. Because attributes can be used as leading indicators of construction safety performance (Tixier et al. 2016b, Esmaeili et al. 2015b), they are also called injury precursors. In what follows, we will use the terms attribute and precursor interchangeably.

Drawing from national databases, Esmaeili and Hallowell (2012, 2011) initially identified 14 and 34 fundamental attributes from 105 fall and 300 struck-by high severity injury cases, respectively. In this study we used a refined and broadened list of 80 attributes carefully engineered and validated by Prades Villanova (2014) and Desvignes (2014) from analyzing a large database of 2,201 reports featuring all injury types and severity levels. These attributes, along with their counts and final risk values in our dataset, are summarized in Table 1. Note that as will be explained later, risk values are unitless and do not have physical meaning. They are only meaningful in that they allow comparison between attributes.

A total of 107 out of 921 reports were discarded because they were not associated with any attribute and because the real outcome was unknown, respectively. Additionally, 3 attributes out of 80 (pontoon, soffit, and poor housekeeping) were removed because they did not appear in any report. This gave a final matrix of $R = 814$ reports by $P = 77$ attributes. While other related studies concerned themselves with pattern recognition and predictive modeling (e.g., Chapters 2 and 3 of the present dissertation, Esmaeili et al. 2015b), here we focus on construction safety risk analysis. The study pipeline is summarized in Figure 1.

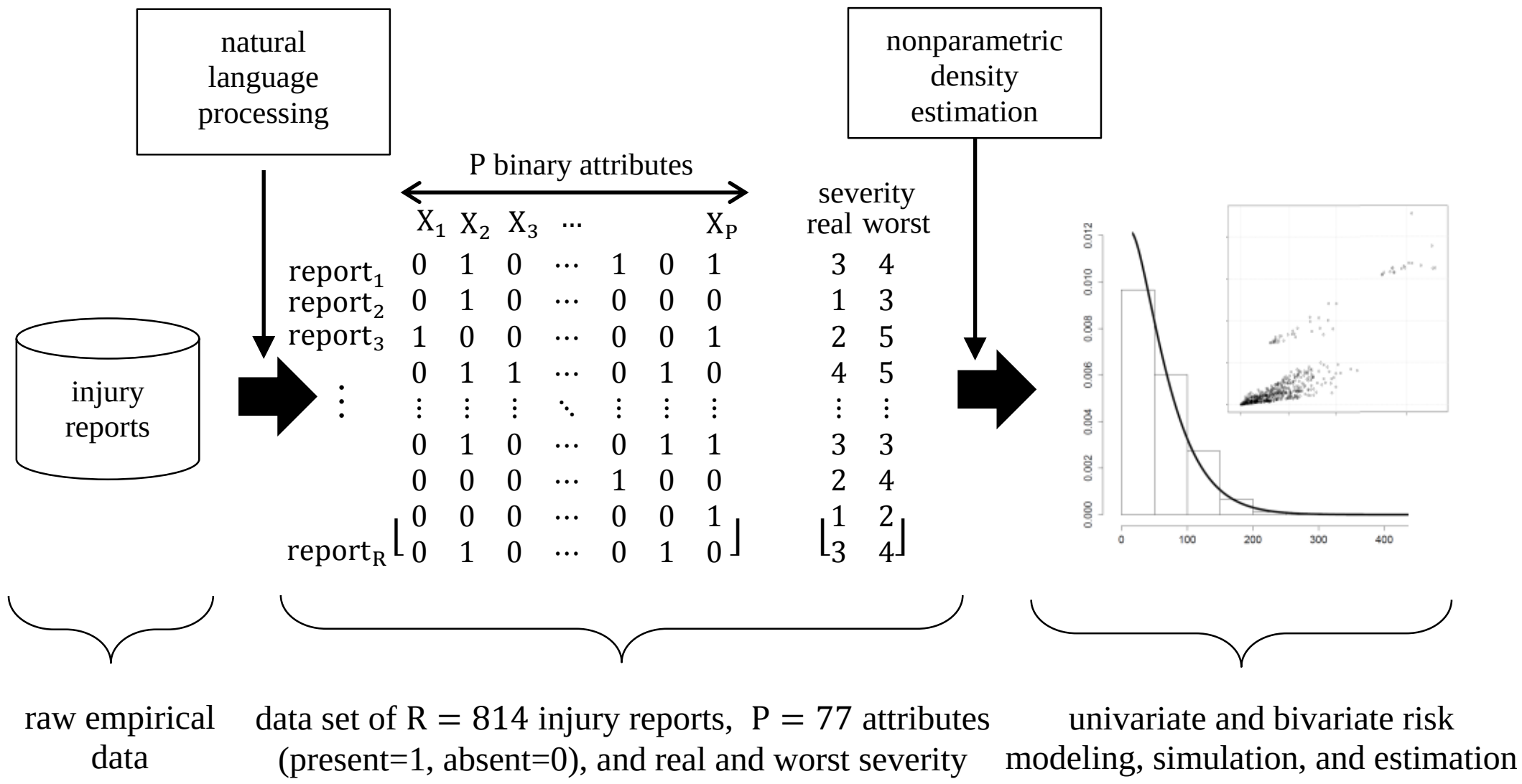

**Figure 1. Overarching research process: from raw injury reports to safety risk analysis**

The contributions of this study are fourfold: (1) we formulate an empirically-grounded definition of construction safety risk at the attribute level, and extend it to the situational level, both in the univariate and the bivariate case; (2) we show how to model risk using Kernel density estimators; (3) we observe that the frequency-magnitude distribution of risk is heavy-tailed, and resembles that of many natural phenomena; and finally, (4) we introduce univariate and bivariate nonparametric stochastic generators based on Kernels and Copulas to draw conclusions from much larger samples and better estimate construction safety risk.

## BACKGROUND AND POINT OF DEPARTURE

To understand how the present study departs from and contributes to the current body of knowledge, we present in what follows a broad review of the safety risk analysis literature. Traditional risk analysis methods for construction safety are limited in two major ways: in terms of the (1) data used (primarily opinion-based), and in terms of the (2) level of analysis (typically trade, activity or task).

**Table 1. Relative risks and counts of the $P = 77$ injury precursors**

| precursor | n | e (%) | risk based on real outcomes | risk based on worst possible outcomes | precursor | n | e (%) | risk based on real outcomes | risk based on worst possible outcomes |
|---|---|---|---|---|---|---|---|---|---|
| concrete | 29 | 41 | 7 | 96 | unstable support/surface | 3 | 32 | 1 | 2 |
| confined workspace | 21 | 2 | 115 | 336 | wind | 29 | 37 | 6 | 16 |
| crane | 16 | 12 | 22 | 76 | improper body position | 7 | 25 | 3 | 6 |
| door | 17 | 21 | 11 | 174 | imp. procedure/inattention | 13 | 16 | 10 | 44 |
| sharp edge | 8 | 38 | 2 | 5 | imp. security of materials | 78 | 12 | 77 | 1007 |
| formwork | 22 | 5 | 63 | 135 | insect | 19 | 18 | 8 | 21 |
| grinding | 16 | 16 | 11 | 34 | no/improper PPE | 3 | 67 | 0* | 1 |
| heat source | 11 | 20 | 4 | 13 | object on the floor | 41 | 43 | 9 | 22 |
| heavy material/tool | 29 | 30 | 11 | 247 | lifting/pulling/handling | 141 | 31 | 49 | 439 |
| heavy vehicle | 12 | 12 | 12 | 307 | cable tray | 9 | 27 | 4 | 11 |
| ladder | 23 | 14 | 15 | 52 | cable | 8 | 33 | 1 | 3 |
| light vehicle | 31 | 59 | 7 | 123 | chipping | 4 | 16 | 1 | 4 |
| lumber | 69 | 14 | 53 | 158 | concrete liquid | 8 | 41 | 2 | 4 |
| machinery | 40 | 8 | 67 | 3159 | conduit | 11 | 31 | 4 | 14 |
| manlift | 8 | 8 | 16 | 50 | congested workspace | 2 | 32 | 0* | 1 |
| object at height | 14 | 50 | 4 | 136 | dunnage | 2 | 16 | 1 | 3 |
| piping | 74 | 38 | 19 | 141 | grout | 3 | 41 | 1 | 1 |
| scaffold | 91 | 33 | 28 | 74 | guardrail handrail | 16 | 40 | 4 | 8 |
| stairs | 28 | 41 | 8 | 25 | job trailer | 2 | 59 | 0* | 1 |
| steel/steel sections | 112 | 35 | 33 | 281 | stud | 4 | 41 | 1 | 5 |
| rebar | 33 | 4 | 76 | 251 | spool | 9 | 33 | 2 | 9 |
| unpowered transporter | 13 | 9 | 23 | 401 | stripping | 12 | 22 | 7 | 18 |
| valve | 24 | 27 | 9 | 22 | tank | 16 | 31 | 5 | 115 |
| welding | 25 | 22 | 10 | 34 | drill | 16 | 43 | 5 | 88 |
| wire | 30 | 43 | 5 | 19 | bolt | 36 | 41 | 7 | 27 |
| working at height | 73 | 40 | 18 | 46 | cleaning | 22 | 56 | 5 | 12 |
| wkg below elev. wksp/mat. | 7 | 17 | 3 | 21 | hammer | 33 | 50 | 5 | 18 |
| forklift | 11 | 9 | 9 | 380 | hose | 11 | 41 | 3 | 8 |
| hand size pieces | 38 | 47 | 7 | 95 | nail | 15 | 50 | 4 | 10 |
| hazardous substance | 33 | 1 | 590 | 6648 | screw | 7 | 50 | 1 | 2 |
| adverse low temps | 33 | 3 | 101 | 292 | slag | 10 | 10 | 8 | 32 |
| mud | 6 | 6 | 9 | 20 | spark | 1 | 12 | 2 | 11 |
| poor visibility | 3 | 23 | 2 | 3 | wrench | 23 | 39 | 5 | 23 |
| powered tool | 32 | 27 | 12 | 54 | exiting/transitioning | 25 | 49 | 6 | 17 |
| slippery surface | 32 | 25 | 13 | 40 | splinter/sliver | 9 | 44 | 1 | 2 |
| small particle | 96 | 31 | 28 | 105 | working overhead | 5 | 40 | 1 | 3 |
| unpowered tool | 102 | 44 | 24 | 352 | repetitive motion | 2 | 51 | 0* | 1 |
| electricity | 1 | 33 | 0* | 1 | imp. security of tools | 24 | 22 | 12 | 314 |
| uneven surface | 33 | 32 | 11 | 129 | | | | | |

* values are rounded up to the nearest integer

## Data

While the data used differ widely among construction safety risk studies, three main sources emerge from the literature: expert opinion, government statistics, and empirical data obtained from construction organizations or national databases. The vast majority of studies use opinion-based data, and thus rely on the ability of experts to rate the relative magnitude of risk based on their professional experience. Often, ranges are provided by researchers to bound risk values. Additionally, even the most experienced experts have limited personal history with hazardous situations, and their judgment under uncertainty suffer the same cognitive limitations as that of any other human being (Capen 1976). Some of these judgmental biases include overconfidence, anchoring, availability, representativeness, unrecognized limits, motivation, and conservatism (Rose 1987, Tversky and Kahneman 1981, Capen 1976). It has also been suggested that gender and even current emotional state have an impact on risk perception (Tixier et al. 2014, Gustafsod 1998). Even if it is possible to somewhat alleviate the negative impact of adverse psychological factors (e.g., Hallowell and Gambatese 2009b), the reliability of data obtained from expert opinion is questionable. Conversely, truly objective empirical data, like the injury reports used in this study, seem superior.

**Level of analysis**

Due to the technological and organizational complexity of construction work, most safety risk studies assume that construction processes can be decomposed into smaller parts (Lingard 2013). Such decomposition allows researchers to model risk for a variety of units of analysis. For example, Hallowell and Gambatese (2009a), Navon and Kolton (2006), and Huang and Hinze (2003) focused on specific tasks and activities. Most commonly, trade-level risk analysis has been adopted (Baradan and Usmen 2006, Jannadi and Almishari 2003, Everett 1999). The major limitation of these segmented approaches is that because each one considers a trade, task, or activity in isolation, it is impossible for the user to comprehensively characterize onsite risk in a standard, robust and consistent way.

Some studies attempted to address the aforementioned limitations. For instance, Shapira and Lyachin (2009) quantified risks for very generic factors related to tower cranes such as type of load or visibility, thereby allowing safety risk modeling for any crane situation. Esmaeili and Hallowell (2012, 2011) went a step further by introducing a novel conceptual framework allowing *any* construction situation to be fully and objectively described by a unique combination of fundamental context-free attributes of the work environment. This attribute-based approach is powerful in that it shows possible the extraction of structured standard information from naturally occurring, unstructured textual injury reports. Additionally, the universality of attributes allows to capture the multifactorial nature of safety risk in the same unified way for any task, trade, or activity, which is a significant improvement over traditional segmented studies. However, manual content analysis of reports is expensive and fraught with data consistency issues. For this reason, Tixier et al. (2016a) introduced a Natural Language Processing (NLP) system capable of automatically detecting the attributes presented in Table 1 and various safety outcomes in injury reports with more than 95% accuracy (comparable to human performance), enabling the large scale use of the attribute-based framework. The data we used in this study was extracted by the aforementioned NLP tool.

## UNIVARIATE ANALYSIS

### Attribute-level safety risk

Following Baradan and Usmen (2006), we defined construction safety risk as the product of frequency and severity as shown in equation 1. More precisely, in our approach, the safety risk $\mathrm{R_p}$ accounted for by $\mathrm{precursor_p}$ (or $\mathrm{X_P}$ in Tables 1 and 2) was computed as the product of the number $\mathrm{n_{ps}}$ of injuries attributed to $\mathrm{precursor_p}$ for the severity level s (given by Table 2) and the impact rating $\mathrm{S_s}$ of this severity level (given by Table 3, and based on Hallowell and Gambatese 2009a). We considered five severity levels, $\mathrm{s_1}$= Pain, $\mathrm{s_2}$= First Aid, $\mathrm{s_3}$= Medical Case/Lost Work Time, $\mathrm{s_4}$= Permanent Disablement, and $\mathrm{s_5}$= Fatality. Medical Case and Lost Work Time were merged because differentiating between these two severity levels turned out to be challenging based on the information available in the narratives only.

$$\mathrm{risk} = \mathrm{frequency} \cdot \mathrm{severity}$$

**Equation 1. Construction safety risk**

**Table 2. Counts of injury severity levels accounted for by each precursor**

| Precursors | Severity levels | | | | |
|---|---|---|---|---|---|
| | $s_1$ = Pain | $s_2$ = 1st Aid | $s_3$ = Medical Case/Lost Work Time | $s_4$ = Permanent Disablement | $s_5$ = Fatality |
| $X_1$ | $n_{11}$ | $n_{12}$ | $n_{13}$ | $n_{14}$ | $n_{15}$ |
| $X_2$ | $n_{21}$ | $n_{22}$ | $n_{23}$ | $n_{24}$ | $n_{25}$ |
| : | : | : | : | : | : |
| $X_{P-1}$ | $n_{(P-1)1}$ | $n_{(P-1)2}$ | $n_{(P-1)3}$ | $n_{(P-1)4}$ | $n_{(P-1)5}$ |
| $X_P$ | $n_{P1}$ | $n_{P2}$ | $n_{P3}$ | $n_{P4}$ | $n_{P5}$ |

**Table 3. Severity level impact scores**

| Severity Level ($s$) | Severity scores ($S_s$) |
|---|---|
| Pain | $S_1 = 12$ |
| 1st Aid | $S_2 = 48$ |
| Medical Case/Lost Work Time | $S_3 = 192$ |
| Permanent Disablement | $S_4 = 1024$ |
| Fatality | $S_5 = 26214$ |

The total amount of risk that can be attributed to $\mathrm{precursor}_p$ was then obtained by summing the risk values attributed to this precursor across all severity levels, as shown in equation 2.

$$R_p = \sum_{s=1}^{5} (n_{ps} \cdot S_s)$$

**Equation 2. Total amount of risk associated with $\mathrm{precursor}_p$**

Where $n_{ps}$ is the number of injuries of severity level s attributed to $\mathrm{precursor}_p$, and $S_s$ is the impact score of severity level s

Finally, as noted by Sacks et al. (2009), risk analysis is inadequate if the likelihood of worker exposure to specific hazards is not taken into account. Hence, the risk $R_p$ of $\mathrm{precursor}_p$ was weighted by its probability of occurrence $e_p$ onsite (see equation 3), which gave the relative risk $RR_p$ of $\mathrm{precursor}_p$. The probabilities $e_p$, or exposure values, were provided by the same company that donated the injury reports. These data are constantly being recorded by means of observation as part of the firm's project control and work characterization policy, and therefore were already available.

$$RR_p = \frac{1}{e_p} \cdot R_p = \frac{1}{e_p} \cdot \sum_{s=1}^{5} (n_{ps} \cdot S_s)$$

**Equation 3. Relative risk for $\mathrm{precursor}_p$**

Where $R_p$ is the total amount of risk associated with $\mathrm{precursor}_p$, and $e_p$ is the probability of occurrence of $\mathrm{precursor}_p$ onsite.

To illustrate the notion of relative risk, assume that the precursor $\mathrm{lumber}$ has caused 15 first aid injuries, 10 medical cases and lost work time injuries, and has once caused a permanent disablement. By following the steps outlined above, the total amount of risk $R_{\mathrm{lumber}}$ accounted for by the attribute lumber can be computed as $15 \times 48 + 10 \times 192 + 1 \times 1024 = 3664$. Moreover, if lumber is encountered frequently onsite, e.g., with an exposure value $e_{\mathrm{lumber}} = 0.65$, the relative risk of lumber will be $RR_{\mathrm{lumber}} = 3664/0.65 = 5637$. On the other hand, if workers are very seldom exposed to $\mathrm{lumber}$ (e.g., $e_{\mathrm{lumber}} = 0.07$), $RR_{\mathrm{lumber}}$ will be equal to $3664/0.07 = 52343$. It is clear from this example that if two attributes have the same total risk value, the attribute having the lowest exposure value will be associated with the greatest relative risk. The assumption is that if a rare attribute causes as much damage as a more common one, the rare attribute should be considered riskier. Note that relative risk values allow comparison but do not have an absolute physical meaning. As presented later, what matters more than the precise risk value itself is the range in which the value falls.

Also, note that since Tixier et al.'s (2016a) NLP tool's functionality did not include injury severity extraction at the time of writing, we used the real and worst possible outcomes manually assessed for each report by Prades Villanova (2014). Specifically, in Prades Villanova (2014), a team of 7 researchers analyzed a large database of injury reports over the course of several weeks. High output quality was ensured by using a harsh 95% inter-coder agreement threshold, peer-reviews, calibration meetings, and random verifications by an external reviewer. Regarding worst possible injury severity, human coders were asked to use their judgment of what would have happened in the worst case scenario should a small translation in time and/or space had occurred. This method and the resulting judgments were later validated by Alexander et al. (2015) who showed that the human assessment of maximum possible severity was congruent with the quantity of energy in the situation, which, ultimately, is a reliable predictor of the worst possible outcome.

For instance, in the following excerpt of an injury report: “worker was welding below scaffold and a hammer fell from two levels above and scratched his arm”, the real severity is a first aid. However, by making only a small translation in space, the hammer could have struck the worker in the head, which could have yielded a permanent disablement or even a fatality. Furthermore, coders were asked to favor the most conservative choice; that is, here, permanent disablement. Whenever mental projection was impossible or required some degree of speculation, coders were required to leave the field as blank and the reports were subsequently discarded. As indicated, Alexander et al. (2015) empirically validated these subjective assessments.

By considering severity counts for both real outcomes and worst possible outcomes, we could compute two relative risk values for each of the 77 precursors. These values are listed in Table 1, and were stored in two vectors of length $P = 77$.

For each attribute, we computed the difference between the relative risk based on worst possible outcomes and the relative risk based on actual outcomes. The top 10% attributes for this metric, which can be considered the attributes that have the greatest potential for severity escalation should things go wrong, are hazardous substance ($\Delta$= 6059), machinery (3092), improper security of materials (930), lifting/pulling/manual handling (390), unpowered transporter (378), forklift (371), unpowered tool (328), improper security of tools (302), and heavy vehicle (295). Except lifting/ pulling/manual handling and unpowered tool, all these precursors are directly associated with human error or high energy levels, which corroborates recent findings (Tixier 2015, Alexander et al. 2015, respectively). Furthermore, one could argue that the attributes lifting/pulling/manual handling and unpowered tool are still indirectly related to human error and high energy levels, as the former is often associated with improper body positioning (human factor) while the latter usually designates small and hand held objects (hammer, wrench, screwdriver, etc.) that are prone to falling from height (high energy). Many attributes in Table 1, such as sharp edge, manlift, unstable support/surface, or improper body position, have low risk values because of their rarity in the rather small data set that we used to illustrate our methodology, but this does not incur any loss of generality.

**Report-level safety risk**

By multiplying the $(R, P)$ attribute binary matrix (attribute matrix of Figure 1) by each $(P, 1)$ relative risk vector (real and worst) as shown in equation 4, two risk values were obtained for each of the $R = 814$ incident reports. This operation was equivalent to summing the risk values based on real and worst possible outcomes of all the attributes that were identified as present in each report (see equation 5).

For instance, in the following description of a construction situation: “worker is unloading a ladder from pickup truck with bad posture”, four attributes are present: namely (1) *ladder*, (2) *manual handling*, (3) *light vehicle,* and (4) *improper body positioning*. The risk based on real outcomes for this construction situation can be computed as the sum of the relative risk values of the four attributes present (given by Table 1), that is, $15 + 49 + 7 + 3 = 74$, and similarly, the risk based on worst potential outcomes can be computed as $52 + 439 + 123 + 6 = 620$.

$$\underset{\text{R reports}}{\updownarrow}\overset{\text{P precursors}}{\begin{bmatrix} 0 & 1 & 0 & \cdots & 1 & 0 & 1 \\ 0 & 1 & 0 & \cdots & 1 & 0 & 0 \\ 1 & 0 & 0 & \cdots & 0 & 0 & 1 \\ 0 & 1 & 1 & \cdots & 0 & 1 & 0 \\ \vdots & \vdots & \vdots & \ddots & \vdots & \vdots & \vdots \\ 0 & 1 & 0 & \cdots & 0 & 1 & 1 \\ 0 & 0 & 0 & \cdots & 0 & 0 & 0 \\ 0 & 0 & 0 & \cdots & 0 & 0 & 1 \\ 0 & 1 & 0 & \cdots & 0 & 1 & 0 \end{bmatrix}_{(R,P)}} \cdot \begin{bmatrix} RR_1 \\ RR_2 \\ RR_3 \\ \vdots \\ RR_{(P-2)} \\ RR_{(P-1)} \\ RR_P \end{bmatrix}_{(P,1)} = \begin{bmatrix} R_{report_1} \\ R_{report_2} \\ R_{report_3} \\ R_{report_4} \\ \vdots \\ R_{report_{(R-3)}} \\ R_{report_{(R-2)}} \\ R_{report_{(R-1)}} \\ R_{report_R} \end{bmatrix}_{(R,1)}$$

**Equation 4. Safety risk at the report level (a)**

Multiplying the $(R, P)$ attribute matrix by the $(P, 1)$ vector of relative risk values for each attribute gives the $(R, 1)$ vector of risk values associated with each injury report.

$$R_{report_r} = \sum_{p=1}^{P} \left( RR_p \cdot \delta_{rp} \right)$$

**Equation 5. Safety risk at the report level (b)**

Where $RR_p$ is the relative risk associated with $precursor_p$, and $\delta_{rp} = 1$ if $precursor_p$ is present in $report_r$ ($\delta_{rp} = 0$ else).

As already stressed, these relative values are not meaningful in absolute terms, they only enable comparison between situations and their categorization into broad ranges of riskiness (e.g., low, medium, high). Estimating these ranges on a small, finite sample such as the one we used in this study would have resulted in biased estimates. To alleviate this, we used stochastic simulation techniques to generate hundreds of thousands of new scenarios honoring the historical data, enabling us to make inferences from a much richer, yet faithful sample.

**The probability distribution of construction safety risk resembles that of many natural phenomena**

For a given injury report, the risk based on real outcomes and the risk based on worst potential outcomes can each take on a quasi-infinite number of values $(2^P - 1)$ with some associated probabilities.

Therefore, they can be considered quasi-continuous random variables, and have legitimate probability distribution functions (PDFs). Furthermore, since a risk value cannot be negative by definition, these PDFs have $[0, +\infty)$ support.

The empirical PDF of the risk based on real outcomes for the 814 injury reports is shown as a histogram in Figure 2. The histogram divides the sample space into a number of intervals and simply counts how many observations fall into each range. We can clearly see that the empirical safety risk is rightly skewed and exhibits a thick tail feature. In other words, the bulk of construction situations present risk values in the small-medium range, while only a few construction situations are associated with high and extreme risk. This makes intuitive sense and is in accordance with what we observe onsite, i.e., frequent benign injuries, and low-frequency high-impact accidents.

Such heavy-tailed distributions are referred to as "power laws" in the literature, after Pareto (1896), who proposed that the relative number of individuals with an annual income larger than a certain threshold was proportional to a power of this threshold. Power laws are ubiquitous in nature (Pinto et al. 2012, Malamud 2004). Some examples of natural phenomena whose magnitude follow power laws include earthquakes, ocean waves, volcanic eruptions, asteroid impacts, tornadoes, forest fires, floods, solar flares, landslides, and rainfall (Papalexiou et al. 2013, Pinto et al. 2012, Menéndez et al. 2008, Malamud et al. 2006). Other human related examples include insurance losses and healthcare expenditures (Ahn et al. 2012), hurricane damage cost (Jagger et al. 2008, Katz 2002), and the size of human settlements and files transferred on the web (Reed 2001, Crovella and Bestavros 1995).

To highlight the resemblance between construction safety risk and some of the aforementioned natural phenomena, we selected four datasets that are standard in the field of extreme value analysis, and freely available from the "extRemes" R package (Gilleland and Katz 2011). We overlaid the corresponding

PDFs with that of construction safety risk. For ease of comparison, variables were first rescaled as shown in equation 6. In what follows, each data set is briefly presented.

$$Z = \frac{X - \min(X)}{\max(X) - \min(X)}$$

**Equation 6. Variable rescaling**

Where X is the variable in the original space and Z is the variable in the rescaled space.

*Summer maximum temperatures in Arizona*

The first dataset reported summer maximum temperatures in Phoenix, AZ, from 1948 to 1990, measured at Sky harbor airport. The observations were multiplied by -1 (flipped horizontally) before rescaling. The distribution is named “max temperature” in Figure 3.

*Hurricane economic damage*

The second dataset (“hurricane damage” in Figure 3) consisted in total economic damage caused by every hurricane making landfall in the United States between 1925 and 1995, expressed in 1995 U.S. $ billion. Following Katz’s (2002) recommendation, all individual storms costing less than $0.01 billion were removed to minimize potential biases in the recording process. The final number of hurricanes taken into account was 86.

*Potomac River peak flow*

The third data set included in our comparison was observations of Potomac River peak stream flow measured in cubic feet per second at Point Rocks, MD, from 1895 to 2000. The observations were divided by $10^5$ before rescaling. The curve is labeled “peak flow” in Figure 3.

*Precipitation in Fort Collins, CO*

The fourth and last dataset contained 36,524 daily precipitation amounts (in inches) from a single rain gauge in Fort Collins, CO. Only values greater than 1 inch were taken into account, giving a final number of 213 observations. The distribution is named “precipitation” in Figure 3.

We estimated the PDFs by using kernel density estimators (KDE) since overlaying histograms would have resulted in an incomprehensible figure. The KDE, sometimes called Parzen, is a nonparametric way to estimate a PDF. It can be viewed as a smoothed version of the histogram, where a continuous function, called the Kernel, is used rather than a box as the fundamental constituent (Silvermann 1986, p. 3). The Kernel has zero mean, is symmetric, positive, and integrates to one. The last two properties ensure that the Kernel, and as a result the KDE, is a probability distribution. More precisely, as shown in equation 7, the KDE at each point x is the average contribution from each of the Kernels at that point (Hastie et al. 2009 p. 208). Put differently, the KDE at x is a local average of functions assigning weights to the neighboring observations $x_i$ that decrease as $|x_i - x|$ increases (Saporta 2011, p. 323, Moon et al. 1995). The “local” estimation is the key feature of this method in enabling to capture the features present in the data. KDEs converge faster to the underlying density than the histogram, and are robust to the choice of the origin of the intervals (Moon et al. 1995).

**Figure 2. Histogram of original observations (n=814) with boundary corrected KDE of the simulated observations (n=10^5)**

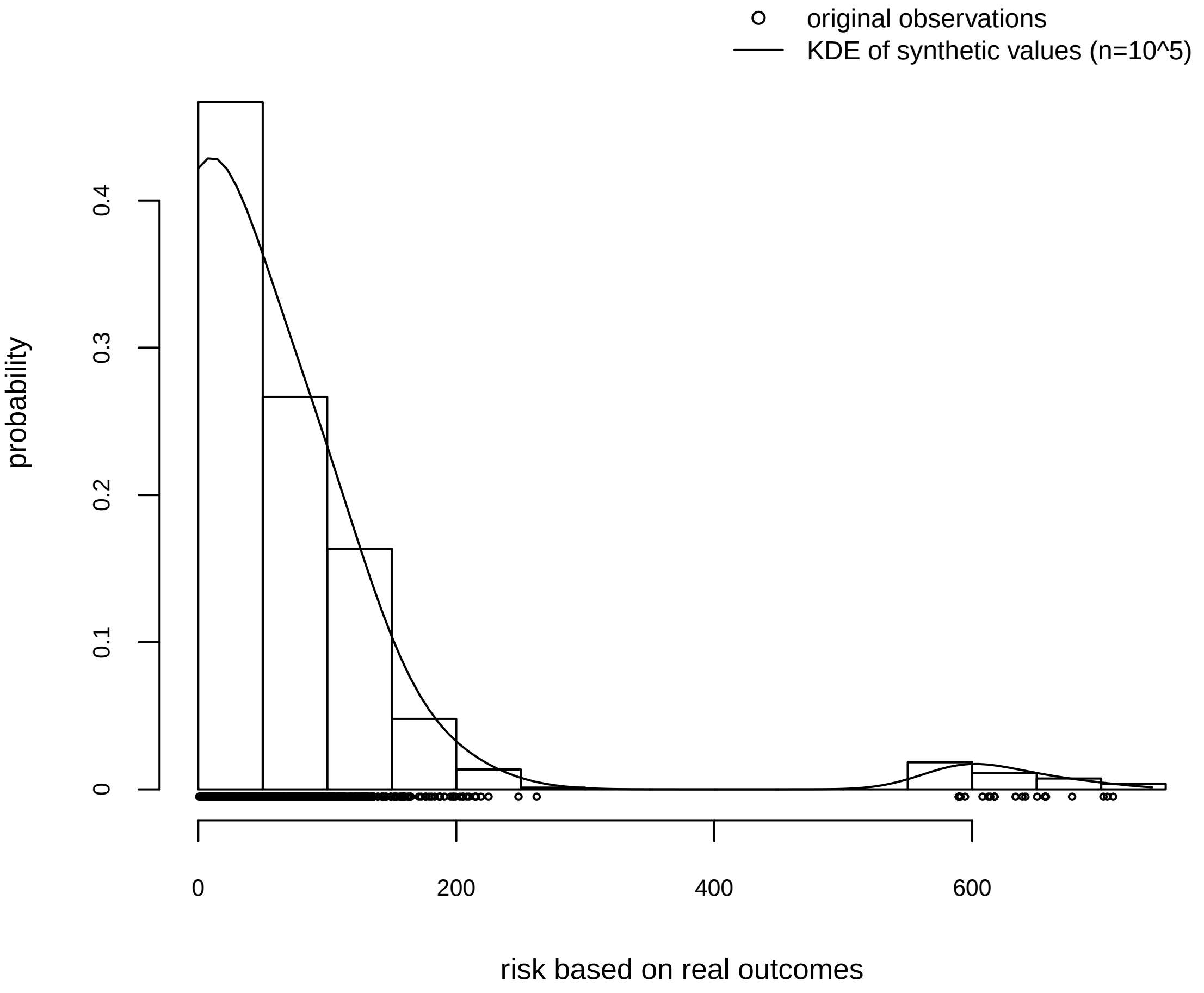

**Figure 3. Comparison of safety risk with natural phenomena**

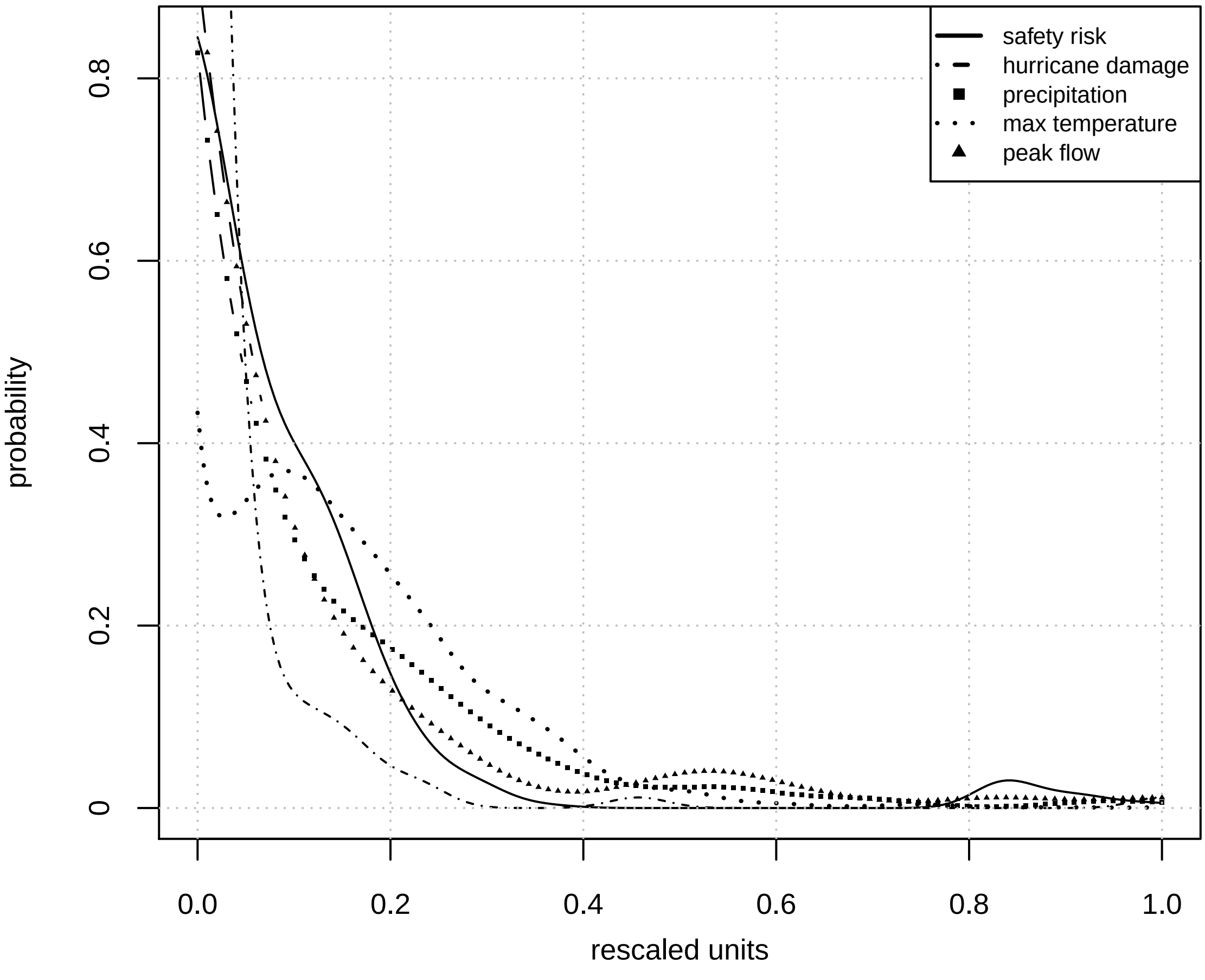

$$\widehat{f_X}(x) = \frac{1}{nh}\sum_{i=1}^{n} K(\frac{x - x_i}{h}, h)$$

**Equation 7. Kernel Density Estimator (KDE)**

Where $\{x_1, \dots, x_n\}$ are the observations, K is the Kernel, and h is a parameter called the bandwidth. Note that $\widehat{f_X}$ is an estimator of the true PDF $f_X$, which is unknown.

h is a parameter called the bandwidth that controls smoothing and therefore affects the final shape of the estimate (Hastie et al. 2009, p. 193). A large bandwidth creates a great amount of smoothing, which decreases variance and increases bias as the fit to the observations is loose. In that case, most of the structure in the data is not captured (i.e., underfitting). On the other hand, a small bandwidth will tightly fit the data and its spurious features such as noise (i.e., overfitting), which yields a low bias but a high variance. There is definitely a tradeoff here. In this study, we used a standard and widespread way of estimating h called Silverman's rule of thumb (Silverman 1986 p. 48) shown in Equation 8. We invite the reader to reference Rajagopalan et al. (1997a) for a good review of the objective bandwidth selection methods.

$$h = \frac{0.9 \min\left(\widehat{\sigma}_X, \frac{Q_3 - Q_1}{1.34}\right)}{n^{1/5}}$$

**Equation 8. Silverman's rule of thumb for bandwidth selection**

Where Q3 and Q1 are the third and first quartiles (respectively), $\widehat{\sigma}_X$ is the standard deviation of the sample, and n is the size of the sample. Here, $n = R = 814$.

Further, for our Kernel K, we selected the standard Normal distribution $N(0,1)$, that is, the Normal distribution centered on zero with unit variance. Since the PDF of N(0,1) is $\frac{1}{\sqrt{2\pi}}e^{-x^2/2}$, the associated KDE can be written using Equation 7 as shown in Equation 9. Other popular Kernels include the triangular, biweight or Epanechnikov, but the consensus in the statistics literature is that the choice of the Kernel is secondary to the estimation of the bandwidth (e.g., Saporta 2011, p. 323).

$$\widehat{f_X}(x) = \frac{1}{nh\sqrt{2\pi}} \sum_{i=1}^{n} e^{-\frac{1}{2}\left(\frac{x_i - x}{h}\right)^2}$$

**Equation 9. KDE with standard Normal Kernel**

Where $\{x_1, \dots, x_n\}$ are the observations, and h is the bandwidth. Here, $n = R = 814$.

It is well known that the KDE suffers a bias at the edges on bounded supports. Indeed, because the Kernel functions are symmetric, weights are assigned to values outside the support, which causes the density near the edges to be significantly underestimated, and creates a faulty visual representation. In our case, safety risk takes on values in $[0, +\infty)$, so issues arise when approaching zero. We used the correction for the boundary bias via local linear regression (Jones 1993) using the "evmix" package (Hu and Scarott 2014) of the R programming language (R core team 2015). Boundary reflection and log transformation are other popular approaches for controlling boundary bias (Rajagopalan et al. 1997a, Silverman 1986).

**Why does construction safety risk follow a power law?**

The power law behavior of construction safety risk can be explained from a technical standpoint by the "inverse of quantities" mechanism. As Newman (2005) explains, any quantity $X \sim Y^{-\gamma}$ for a given $\gamma$ will have a probability distribution $P[X] \sim X^{-\alpha}$, with $\alpha = 1 + 1/\gamma$. Further, it can be shown that this probability distribution exhibits power law behavior.

In the special case of construction safety risk, by simply using the fact that $RR_p = \frac{1}{e_p} \cdot R_p$ (equation 3), we can rewrite equation 5 as equation 10.

$$R_{report_r} = \sum_{p=1}^{P} \left( \frac{1}{e_p} \cdot \ R_p \ \cdot \ \delta_{rp} \right)$$

**Equation 10. Risk at the report level (c)**

Where $e_p$ is the probability of occurrence of $precursor_p$ onsite, $R_p$ is the total amount of risk associated with $precursor_p$, and $\delta_{rp} = 1$ if $precursor_p$ is present in $report_r$ (0 else).

Finally, setting $X = R_{report_r}$ and $Y = \prod_1^p e_p$ , it follows from equation 10 that $X \sim \ Y^{-\gamma}$ with $\gamma = 1$, which, according to Newman (2005), suffices to show that $R_{report_r}$ is power law distributed. Further, Newman (2005) stresses that even though the relationship between $X$ and $Y$ is already some sort of power law ($X$ is proportional to a power of $Y$), this relationship is deterministic, not stochastic. Still, it generates a power law probability distribution, which is not trivial.

Moreover, the large values of $R_{report_r}$, those in the tail of the distribution, correspond to large values of $RR_p$, that is, to small values of $e_p$ close to zero (i.e., rare precursors). This makes sense, and is in accordance with the theory of extremes (extreme values are rare).

There are more underlying processes that can generate fat tails in the distributions of natural and other human-related phenomena, such as multiplicative processes (Adlouni et al. 2008, Mitzenmacher 2004) random walks, the Yule process, self-organized criticality, and more (Newman 2005). They cannot be all addressed here. Moreover, the *inverse of quantities* mechanism seems to be the most plausible and most straightforward explanation for the shape of the probability distribution of construction safety risk observed in this study.

## Univariate modeling

In this section, we focus on construction safety risk based on real outcomes. We present a computational method that can be used to generate synthetic safety risk values that honor the historical data. Note that while many techniques and concepts in risk modeling and management deal with extreme values only, in this study we seek to capture and simulate from the entire risk spectrum (not only the extremes) in order to accurately assess the safety risk of any construction situation.

Today, extreme value analysis is still a subject of active research, and is widely used in a variety of different fields. In addition to the modeling of extreme hydroclimatological events, its applications include insurance losses (Guillen et al. 2011) and financial market shock modeling (Glantz and Kissell 2014). A central quantity in risk management is the quantile.

The quantile function (or simply quantile, for short) of a continuous random variable X is defined as the inverse of its cumulative distribution function (CDF) as shown in equation 11. The CDF is obtained by integrating or summing the PDF, respectively in the continuous and discrete case.

$$Q(p) = {F_X}^{-1}(p)$$

**Equation 11. Quantile function**

Where $F_X$ is the CDF of X defined as $F_X(x) = P[X \leq x] = p \in [0,1]$

The quantile is closely linked to the concept of exceedances. In finance and insurance for instance, the value-at-risk for a given horizon is the loss that cannot be exceeded with a certain probability of confidence within the time period considered, which is given by the quantile function. For instance, the 99.95% value-at-risk $Q(99.95)$ at 10 days represents the amount of money that the loss can only exceed with 0.5% probability in the next 10 days. In other words, the corresponding fund reserve would cover 199 losses over 200 (199/200=0.995).

The quantile function is also associated with the notion of return period $\mathrm{T}$ in hydroclimatology. For example, the magnitude of the 100-year flood ($\mathrm{T} = 100$) corresponds to the streamflow value that is only exceeded by 1% of the observations, assuming one observation per year. This value is given by $\mathrm{Q}(1 - 1/\mathrm{T}) = \mathrm{Q}(0.99)$, which is the 99$^{th}$ percentile, or the 99$^{th}$ 100-quantile. Similarly, the magnitude of the 500-year flood, $\mathrm{Q}(0.998)$, is only exceeded by 0.2% of the observations. For construction safety, this quantity would correspond to the minimum risk value that is only observed on average in one construction situation over five hundred. The median value, given by $\mathrm{Q}(0.5)$, would correspond to the safety risk observed on average in one construction situation over two.

**Limitations of traditional parametric techniques**

Traditional approaches to quantile estimation are based on parametric models of PDF especially from the Extreme Value Theory (EVT) framework (Coles et al. 2001). These models possess fat tails unlike traditional PDFs, and thus are suitable for robust estimation of extremes. The candidate distributions from the EVT are Frechet, Weibull, Gumbel, GEV, Generalized Pareto, or mixtures of these distributions (Charpentier and Oulidi 2010). These parametric models are powerful in that they allow complex phenomena to be entirely described by a single mathematical equation and a few parameters. However, being parametric, these models tend to be suboptimal when little knowledge is available about the phenomenon studied (which is the case in this exploratory study). Indeed, even when enough data are available and all parameters are estimated accurately, conclusions may be irrelevant in the case of initial model misspecification (Charpentier and Oudini 2010, Charpentier et al. 2007, Breiman 2001a). This is very problematic, especially when risk-based decisions are to be made from these conclusions.

In addition, parametric models, even from the EVT, are often too lightly tailed to avoid underestimating the extreme quantiles (Vrac and naveau 2007), which is a major limitation as accurately capturing the tail of a probability distribution is precisely the crucial thing in risk management (Figressi et al. 2002). A popular remediation strategy consists in fitting a parametric model to the tail only, such as the Generalized Pareto, but selecting a threshold that defines the tail is a highly subjective task (Scarrott and MacDonald 2012), and medium and small values, which represent the bulk of the data are overlooked (Vrac and Naveau 2007). What is clearly better, however, especially when the final goal is simulation, is to capture the entire distribution. As a solution, hydroclimatologists have proposed dynamic mixtures of distributions, based on weighing the contributions of two overlapping models, one targeting the bulk of the observations, and the other orientated towards capturing extremes (Furrer and Katz 2007, Frigessi et al. 2002). Unfortunately, threshold selection implicitly carries over through the estimation of the parameters of the mixing function, and with most mixing functions, conflicts arise between the two distributions around the boundary (Hu and Scarrott 2013). For all these reasons, we decided to adopt a fully data-driven, nonparametric approach that we describe below.

**Univariate construction safety risk generator**

The proposed approach consists in generating independent realizations from the nonparametric PDF estimated via the KDE described above. We base our generator on the smoothed bootstrap with variance correction proposed by Silverman (1986, p. 142-145). Unlike the traditional nonparametric bootstrap (Efron 1979) that simply consists in resampling with replacement, the smoothed bootstrap can generate values outside of the historical limited range, and does not reproduce spurious features of the original data such as noise (Rajagopalan et al. 1997b). The smoothed bootstrap approach has been successfully used in modeling daily precipitation (Lall et al., 1996), streamflow (Sharma et al., 1997) and daily weather (Rajagopalan et al., 1997b).

More precisely, the algorithm that we implemented in R to generate our synthetic values can be broken down into the following steps:

For j in 1 to the desired number of simulated values:

1. choose i uniformly with replacement from $\{1, \dots, \mathrm{R}\}$
2. sample $\epsilon_X$ from the standard normal distribution with variance ${h_X}^2$
3. record $X_sim_j = \bar{X} + (X_i - \bar{X} + \epsilon_X) / \sqrt{1 + {h_X}^2 / {\hat{\sigma}_X}^2}$

Where $R = 814$ is the sample size (the number of injury reports), $\bar{X}$ and ${\hat{\sigma}_X}^2$ are the sample mean and variance, and ${h_X}^2$ is the variance of the standard normal Kernel (bandwidth of the KDE). Note that we deleted the negative simulated values to be consistent with the definition of risk.

Figure 2 shows the KDE of the $10^5$ simulated values overlaid with the histogram of the original sample. It can be clearly seen that the synthetic values are faithful to the original sample since the PDF from the simulated values fit the observations very well. Also, while honoring the historical data, the smoothed bootstrap generated values outside the original limited range, as desired. The maximum risk value in our sample was 709, while the maximum of the simulated values was 740 (rounded to the nearest integer). Table 4 compares the quantile estimated via the quantile() R function of the original and simulated observations.

**Table 4. Quantile estimates based on original and simulated values for the risk based on real outcomes**

| | safety risk observed in one situation over: | | | | | | |
|---|---|---|---|---|---|---|---|
| | 2 | 5 | 10 | 100 | 500 | 1,000 | 10,000 |
| Original observations ($n = R = 814$) | 57 | 110 | 152 | 649 | 703 | 706 | 709 |
| Simulated observations ($n = 10^5$) | 61 | 116 | 154 | 647 | 700 | 708 | 728 |

The quantile estimates of Table 4 are roughly equivalent before reaching the tails. This is because the bulk of the original observations were in the low to medium range, enabling quite accurate quantile estimates for this range in the first place. The problem stemmed from the sparsity of the high to extreme values in the historical sample, which made estimation of the extreme quantiles biased. Our use of the smoothed bootstrap populated the tail space with new observations, yielding a slightly higher estimate of the extreme quantiles, as can be seen in Table 4. It makes sense that the extremes are higher than what could have been inferred based simply on the original sample, as the original sample can be seen as a finite window in time whereas our simulated values correspond to observations that would have been made over a much longer period. The chance of observing extreme events is of course greater over a longer period of time.

Based on estimating the quantiles on the extended time frame represented by the synthetic values, we propose the risk ranges shown in Table 5. As already explained, these ranges are more robust and unbiased as the ones that would have been built from our historical observations. Thanks to this empirical way of assessing safety risk, construction practitioners will be able to adopt an optimal proactive approach by taking coherent preventive actions and provisioning the right amounts of resources.

**Table 5. Proposed ranges for the risk based on real outcomes**

| | | | | | | |
|---|---|---|---|---|---|---|
| quantiles | 0 | 0.25 | 0.50 | 0.75 | 0.99 | 1 |
| risk value | 0 | 29 | 61 | 105 | 647 | 740 |
| range | *low* | *medium* | *high* | *very high* | *extreme* | |

## BIVARIATE ANALYSIS

In what follows, we study the relationship between the risk based on real outcomes (X, for brevity) and the risk based on worst potential outcomes (Y). Rather than assuming that these variables are independent and considering them in separation, we acknowledge their dependence and aim at capturing it, and fatefully reproducing it in our simulation engine. This serves the final goal of being able to accurately assess the potential of an observed construction situation for safety risk escalation should the worst case scenario occur. Figure 4 shows a plot of Y versus X, while a bivariate histogram can be seen in Figure 5.

We can distinguish three distinct regimes in Figure 4. The first regime, corresponding roughly to $0 < \mathrm{X} < 70$, is that of benign situations that stay benign in the worst case. Under this regime, there is limited potential for risk escalation. The second regime ($70 < \mathrm{X} < 300$) shows that beyond a certain threshold, moderately risky situations can give birth to hazardous situations in the worst case. The attribute responsible for the switch into this second regime is $\mathrm{machinery}$ (e.g., welding machine, generator, pump). The last regime ($X > 300$) is that of the extremes, and features clear and strong upper tail dependence. The situations belonging to this regime are hazardous in their essence and create severe outcomes in the worst case scenarios. In other words, those situations are dangerous in the first place and unforgiving. The attribute responsible for this extreme regime is $\mathrm{hazardous\ substance}$ (e.g., corrosives, adhesives, flammables, asphyxiants). Again, note that these examples are provided as a result of applying our methodology on a data set of 814 injury reports for illustration purposes but do not incur any loss of generality. Using other, larger data sets would allow risk regimes to be characterized by different and possibly more complex attribute patterns.

## Figure 4. Bivariate construction safety risk

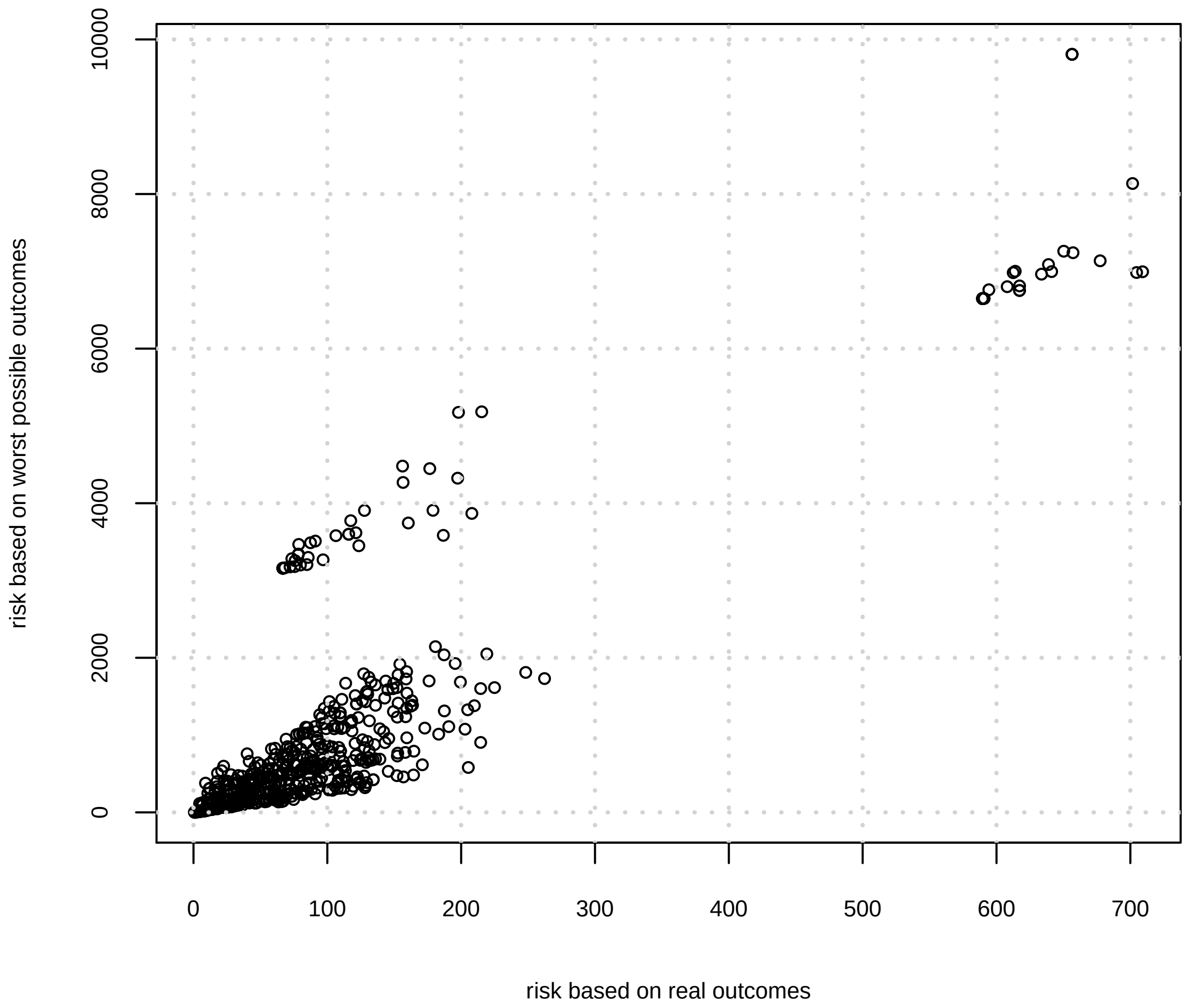

## Figure 5. Bivariate histogram with 25 by 25 grid

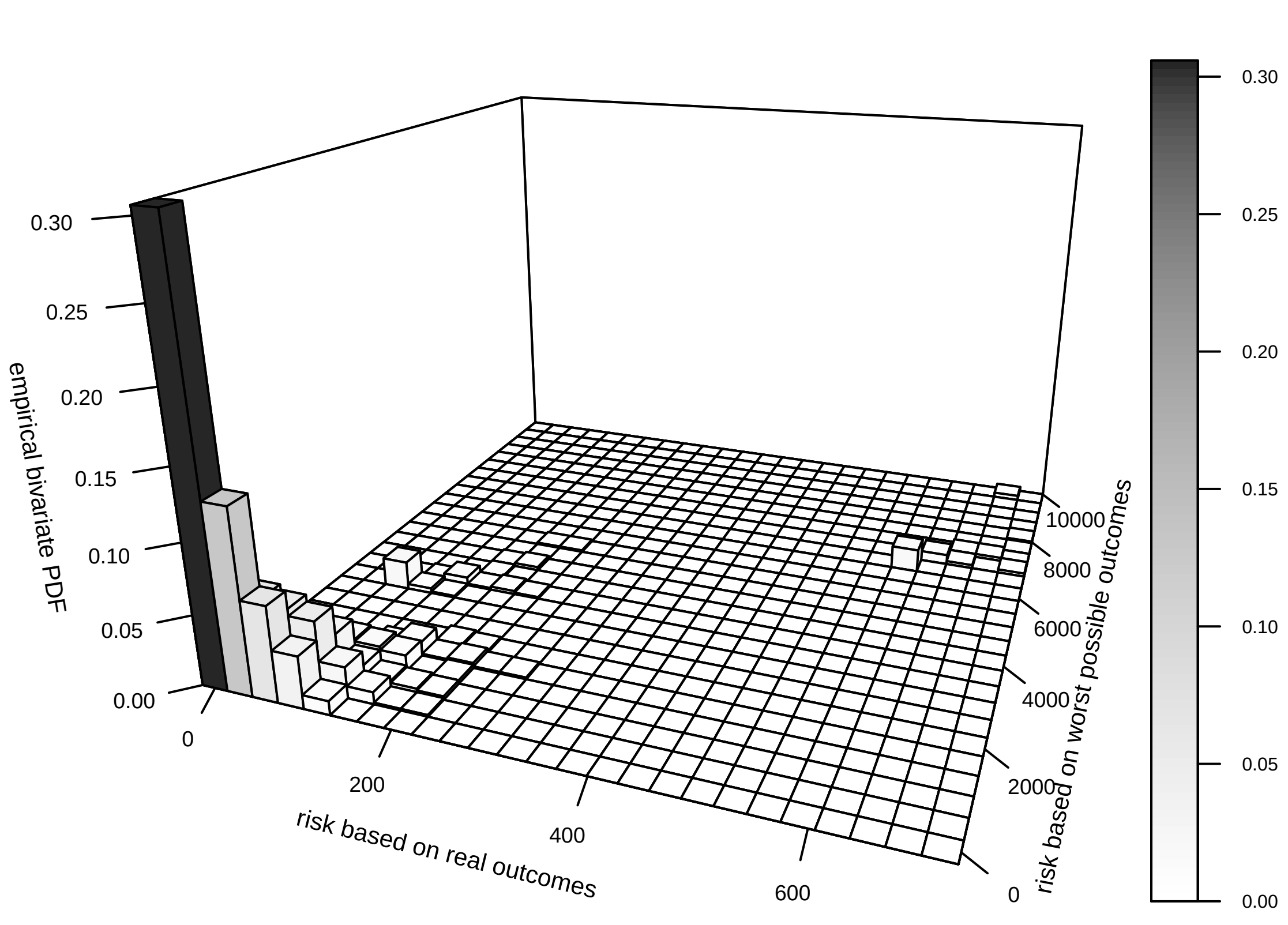

**Bivariate modeling**

Many natural and human-related phenomena are multifactorial in essence and as such their study requires the joint modeling of several random variables. Traditional approaches consist in modeling their dependence with the classical family of multivariate distributions, which is clearly limiting, as it requires all variables to be separately characterized by the same univariate distributions (called the margins). Using Copula theory addresses this limitation by creating a joint probability distribution for two or more variables while preserving their original margins (Hull 2006). In addition to the extra flexibility they offer, the many existing parametric Copula models are also attractive in that they can model the dependence among a potentially very large set of random variables in a parsimonious manner (i.e., with only a few parameters). For an overview of Copulas, one may refer to Cherubini et al. (2004).

While the introduction of Copulas can be tracked back as early as 1959 with the work of Sklar, they did not gain popularity until the end of the 1990s when they became widely used in finance. Copulas are now indispensable to stochastic dependence problem understanding (Durante et al. 2010), and are used in various fields from cosmology to medicine. Since many hydroclimatological phenomena are multidimensional, Copulae are also increasingly used in hydrology, weather and climate research, for instance for precipitation infilling, drought modeling, and extreme storm tide modeling (Bárdossy et al. 2014, Domino et al. 2014, Salvadori et al. 2007).

Formally, a d-dimensional Copula is a joint CDF with $[0,1]^d$ support and standard uniform margins (Charpentier 2006). Another equivalent definition is given by Sklar's (1959) theorem, which states in the bivariate case that the joint CDF $F(x, y)$ of any pair (X,Y) of continuous random variables can be written as in Equation 12.

$$F(x, y) = C\{F_X(x), F_Y(y)\}, \quad (x, y) \in \mathbb{R}^2$$

**Equation 12. Sklar's theorem**

Where $F_X$ and $F_Y$ are the respective margins of $X$ and $Y$, and $C: [0,1]^2 \rightarrow [0,1]$ is a Copula.

Note that equation 12 is consistent with the first definition given, because for any continuous random variable $X$ of CDF $F_X$, $F_X(X)$ follows a uniform distribution.

However, parametric Copulas suffer from all the limitations inherent to parametric modeling briefly evoked previously. Therefore, like in the univariate case, we decided to use a fully data-driven, nonparametric approach to Copula density estimation. We used the bivariate KDE to estimate the empirical Copula, which is defined as the joint CDF of the rank-transformed (or pseudo) observations. The pseudo-observations are obtained as shown in Equation 13.

$$U_X(x) = \frac{\text{rank}(x)}{\text{length}(X) + 1}$$

**Equation 13. Rank-transformation.**

Where $U_X$ is the transformed sample of the pseudo observations,

and $X$ is the original sample.

Because the Copula support is the unit square $[0,1]^2$, the KDE boundary issue arises twice this time, near zero and one, yielding multiplicative biases (Charpentier et al. 2007). Therefore, the density is even more severely underestimated than in the univariate case, and it is even more crucial to ensure robustness of the KDE at the corners to ensure proper visualization. For this purpose, we used the transformation trick described by Charpentier et al. (2007) as our boundary correction technique. The original idea was proposed by Devroye and Györfi (1985). The approach consists in using a transformation $T$ bijective, strictly increasing, continuously differentiable, and which has a continuously differentiable inverse, such that $X' = T(X)$ is unbounded. A KDE can therefore be used to estimate the density of $X'$ without worrying about boundary bias. Finally, a density estimate of $X$ can be obtained via back-transformation, as shown in equation 14.

$$\hat{f}_X(x) = \left.\frac{\hat{f}_{X'}(x')}{\left|\frac{d}{dx'}T^{-1}(x')\right|}\right|_{x'=T(x)}$$

**Equation 14. Transformation trick**

Where $\hat{f}_X$ is the boundary-corrected KDE of X, $\hat{f}_{X'}$ is the KDE of $X'$, and $T$ is the transformation such that

$$X' = T(X)$$

We used the inverse CDF of the Normal distribution, ${F_{N(0,1)}}^{-1}$ as our transformation T. It goes from $[0,1]$ to the real line. The resulting empirical Copula density estimate of the original sample is shown in Figure 6.

**Bivariate construction safety risk generator**

Like in the univariate case, we used a nonparametric, fully data driven approach, the *smoothed bootstrap with variance correction*, as our simulation scheme. Minor adaptations were needed due to the two-dimensional nature of the task. The steps of the algorithm that we implemented using the R programming language are outlined below, and the resulting $10^5$ simulated values are shown in Figure 7. Note that the procedure is equivalent to simulating from the nonparametric Copula density estimate introduced above. Like in the univariate case, we deleted the negative simulated values to ensure consistency with the definition of risk.

For j in 1 to the desired number of simulated values:

1. choose i uniformly with replacement from $\{1, \ldots, R\}$
2. sample $\epsilon_X$ from the standard normal distribution with variance ${h_X}^2$, and $\epsilon_Y$ from the standard normal distribution with variance ${h_Y}^2$
3. take:

$$X_sim_j = \overline{X} + (X_i - \overline{X} + \epsilon_X) \Big/ \sqrt{1 + {h_X}^2 / {\hat{\sigma}_X}^2}$$

$$Y_sim_j = \overline{Y} + (Y_i - \overline{Y} + \epsilon_Y) \Big/ \sqrt{1 + {h_Y}^2 / {\hat{\sigma}_Y}^2}$$

4. record:

$U_sim_j = F_{N(0,1)}(X_sim_j)$, $V_sim_j = F_{N(0,1)}(Y_sim_j)$

Where $R = 814$ is the number of injury reports, $\overline{X}$ and ${\hat{\sigma}_X}^2$ are the mean and variance of X; $\overline{Y}$ and ${\hat{\sigma}_Y}^2$ are the mean and variance of Y; ${h_X}^2$ is the bandwidth of the KDE of X; ${h_Y}^2$ is the bandwidth of the KDE of Y; and $F_{N(0,1)}$ is the CDF of the standard Normal distribution, the inverse of our transformation T.

Note that step 1 selects a pair (x,y) of values from the original sample (X,Y), not two values independently. This is crucial in ensuring that the dependence structure is preserved. Step 4 sends the simulated pair to the pseudo space to enable visual comparison with the empirical Copula density estimate, which is defined in the unit square (i.e., rank space). We can clearly observe in Figure 7 that our sampling scheme was successful in generating values that reproduce the structure present in the original data, validating our nonparametric approach. For the sake of completeness, we also compared (see Figures 8 and 9) the simulated pairs in the original space with the original values. Once again, it is easy to see that the synthetic values honor the historical data. To enable comparison with the univariate case (see Table 4), Table 6 summarizes the empirical quantiles for the historical and simulated observations of risk based on worst potential outcomes (i.e., Y). Like in the univariate case, we can observe that the differences between the estimates increase with the quantiles. Notably, simulation allows to obtain richer estimates of the extreme quantiles, $Q(1-\frac{1}{1000}) = Q(0.999)$ and $Q(1-\frac{1}{10000}) = Q(0.9999)$, whereas with the initial limited sample, the values of the quantile function plateau after $Q(1-\frac{1}{500}) = Q(0.998)$ due to data sparsity in the tail. Similarly to Table 5, we also propose in Table 7 ranges for the risk based on worst potential outcomes.

**Table 6. Quantile estimates based on original and simulated values for the risk based on worst potential outcomes**

| | safety risk observed in one situation over: | | | | | | |
|---|---|---|---|---|---|---|---|
| | 2 | 5 | 10 | 100 | 500 | 1,000 | 10,000 |
| original observations (n = R = 814) | 343 | 950 | 1719 | 7000 | 9808 | 9808 | 9808 |
| simulated observations ($n = 10^5$) | 395 | 1061 | 1953 | 7092 | 9765 | 9586 | 10045 |

**Table 7. Proposed ranges for the risk based on worst potential outcomes**

| | | | | | | |
|---|---|---|---|---|---|---|
| quantiles | 0 | 0.25 | 0.50 | 0.75 | 0.99 | 1 |
| risk value | 0 | 183 | 395 | 837 | 7092 | 10126 |
| range | *low* | *medium* | *high* | *very high* | *extreme* | |

**Figure 6. Nonparametric Copula density estimate with original pseudo−observations**

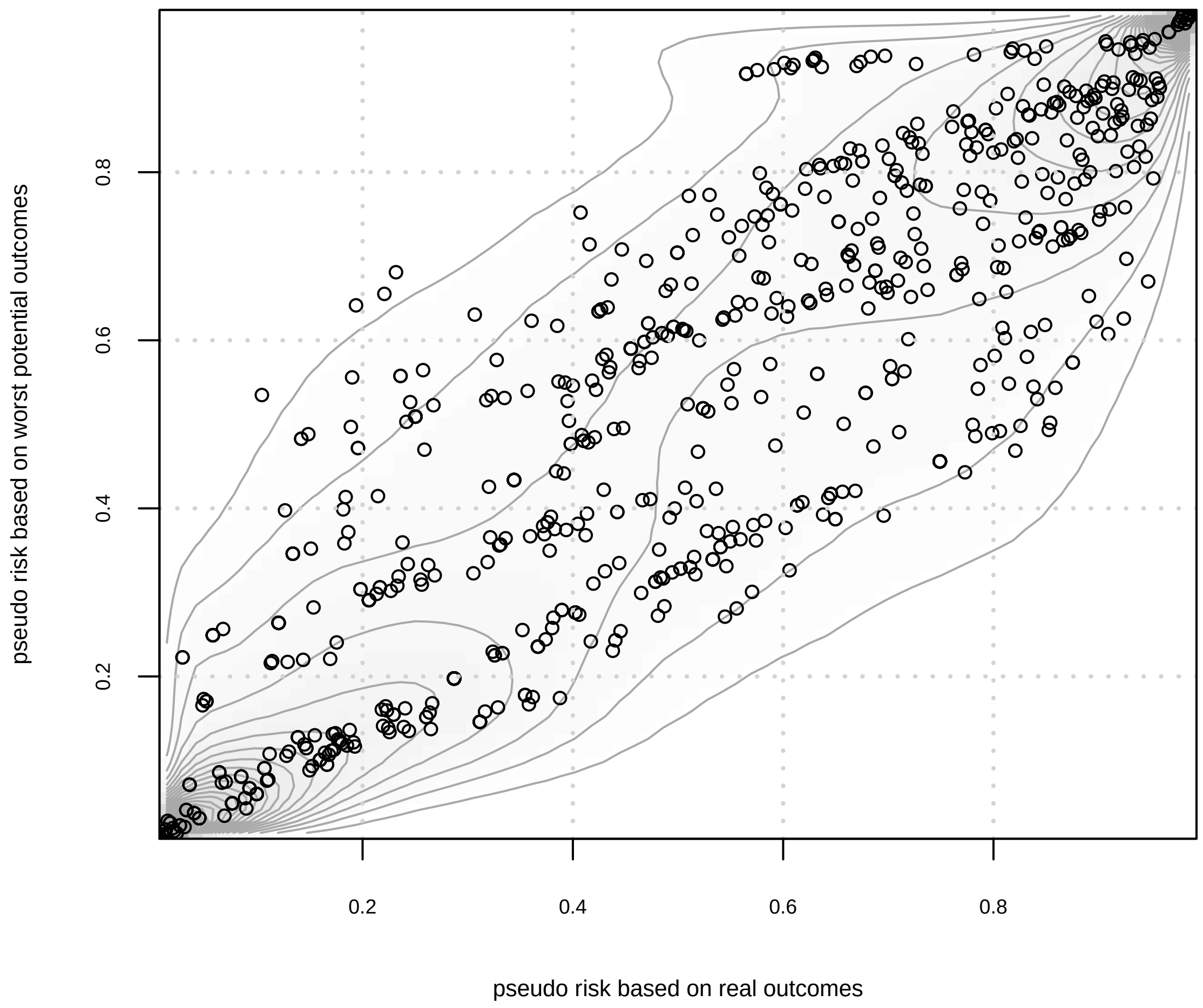

**Figure 7. Simulated risk values in rank space n=10^5**

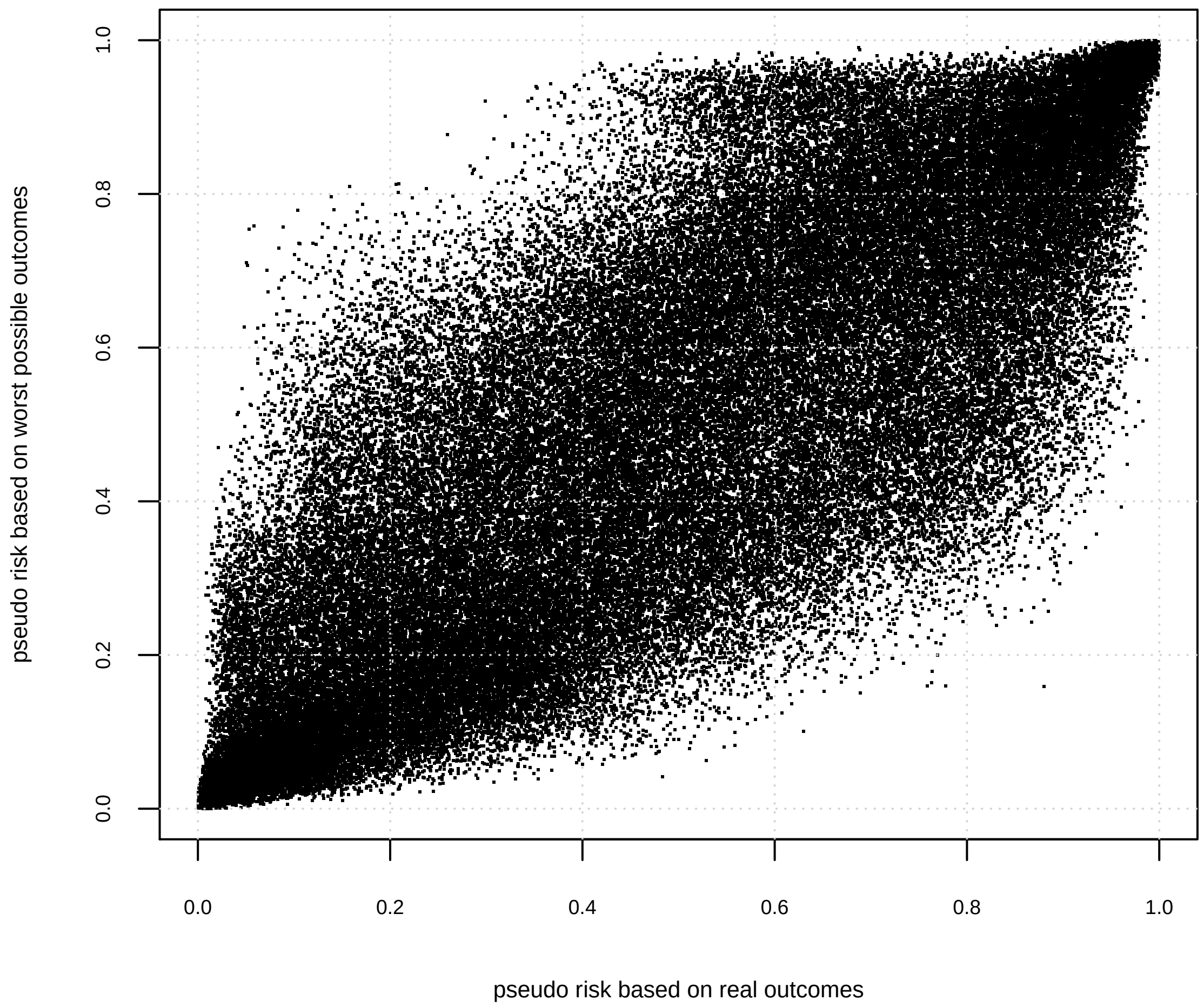

**Figure 8. Bivariate construction safety risk**

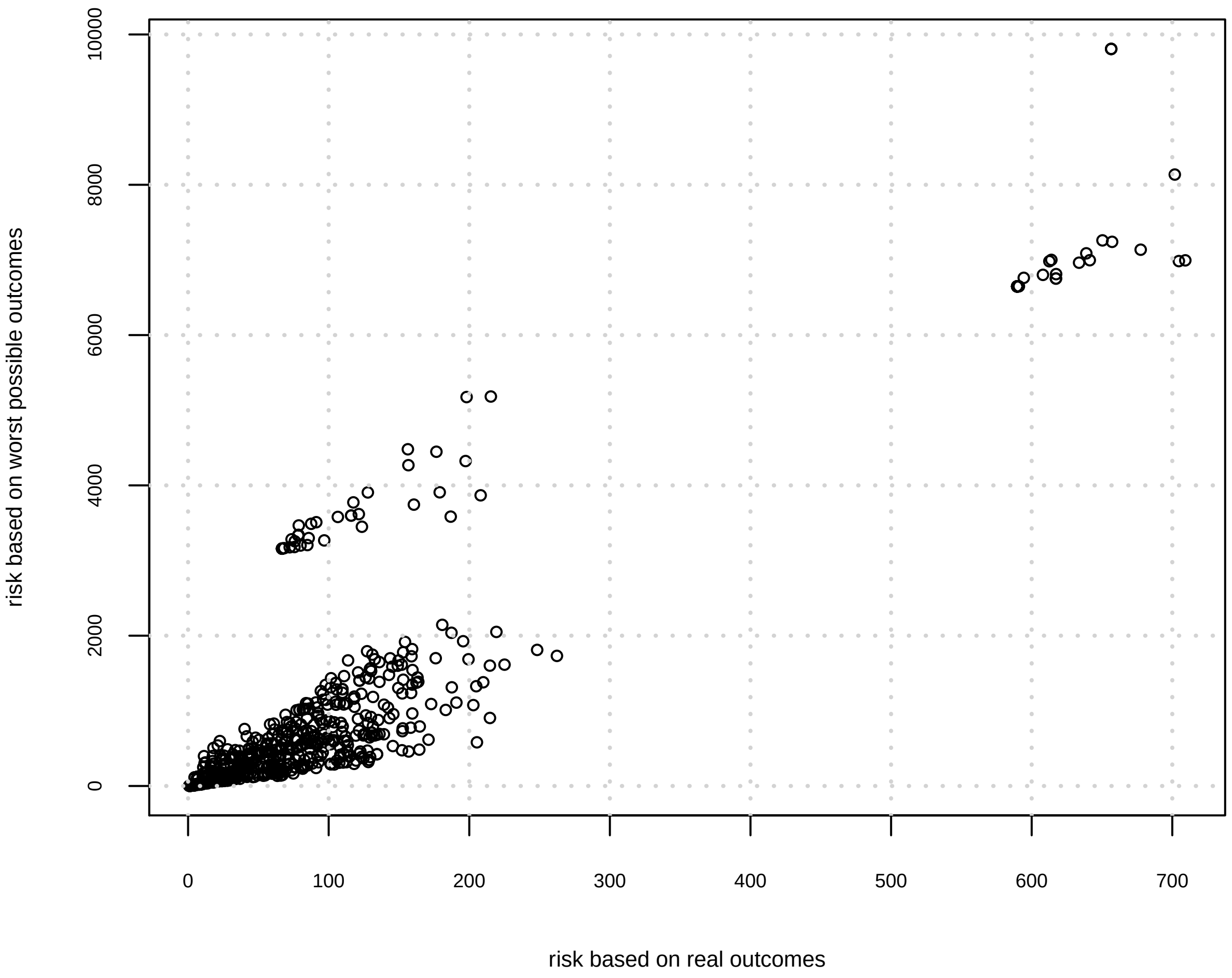

**Figure 9. simulated risk values in original space**
**n=10^5**

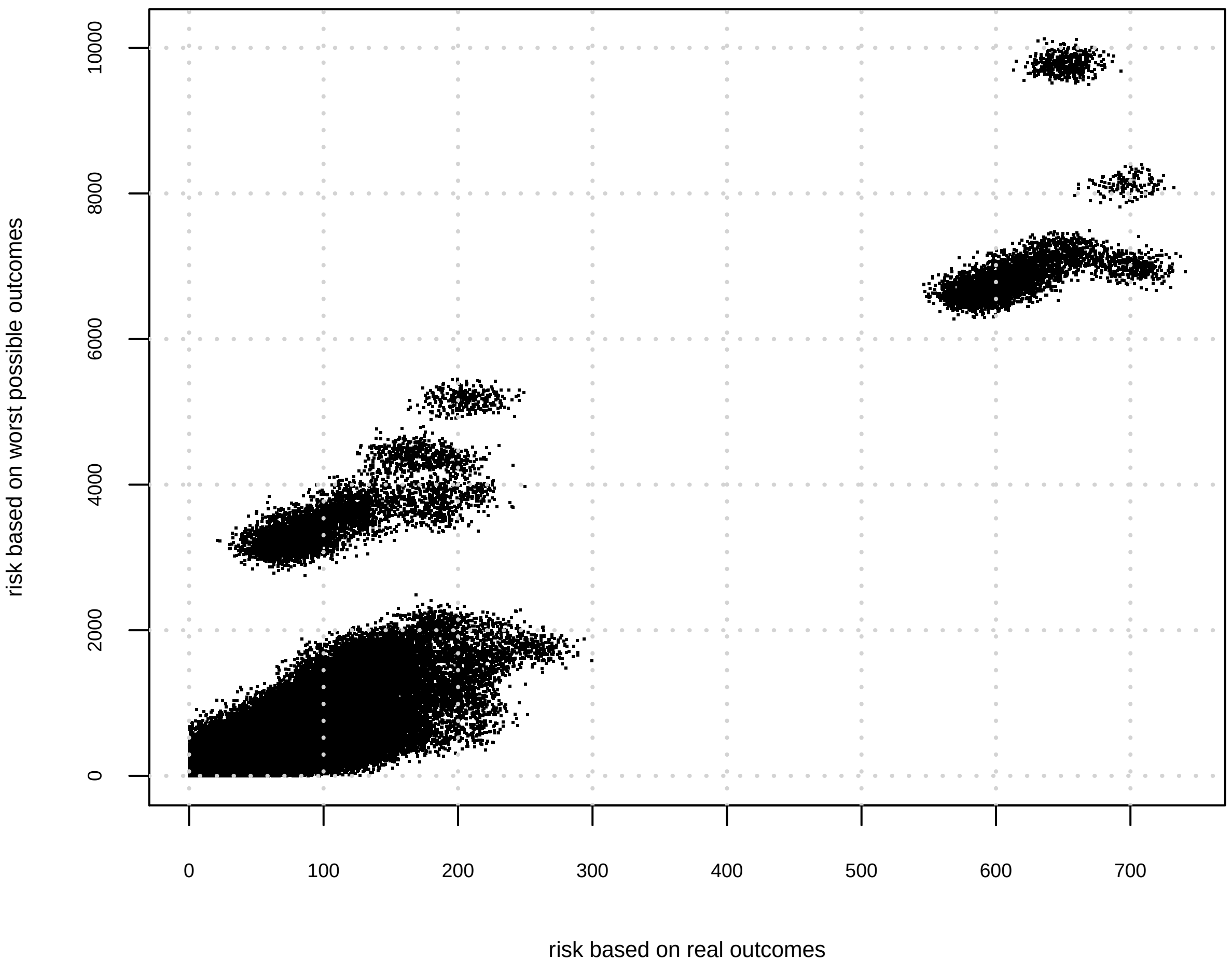

**Computing risk escalation potential based on simulated values**

Using the synthetic safety risk pairs obtained via our bivariate stochastic safety risk generator, and evidence provided by the user (i.e., an observation made onsite in terms of attributes), it is possible to compute and estimate of the upper limit of risk, i.e., the safety risk presented by the observed construction situation based on worst case scenarios. This estimate is based on large numbers of values simulated in a data-driven approach that features the same dependence structure as the original, empirical data. The end user (e.g., designer or a safety manager) can therefore make data-based, informed decisions, and proactively implement the adequate remediation strategies. Furthermore, the attribute-based nature of the procedure is ideally suited for automated integration with building information modeling and work packaging. The technique we propose, based on conditional quantile estimation, consists in the steps detailed in what follows.

First, the attributes observed in a particular construction situation give the risk based on real outcomes for the construction situation, say $x_0$. By fixing the value of X to $x_0$, this first step extracts a slice from the empirical bivariate distribution of the simulated values. This slice corresponds to the empirical probability distribution of Y conditional on the value of X, also noted $P[Y|X = x_0]$. Because only a few values of Y may exactly be associated with $x_0$, we consider all the values of Y associated with the values of X in a small neighboring range around $x_0$, that is, $P[Y|x_0 - x_- < X < x_0 + x_+]$. In our experiments, we used $x_- = x_+ = 5$; that is, a range of $[-5, +5]$ around $x_0$, because it gave good results, but there is no absolute and definitive best range. The second step simply consists in evaluating the quantile function of $P[Y|x_0 - x_- < X < x_0 + x_+]$ at some threshold. The reader can refer to equation 10 for the definition of the quantile function. In our experiments, we used a threshold of 80%, (i.e., we computed $Q(0.8)$ with the quantile() R function), but the choice of the threshold should be made at the discretion of the user, depending on the desired final interpretation. In plain English, the threshold we selected returns the risk based on worst possible outcomes that is only exceeded in 20% of cases for the particular value of risk based on real outcomes computed at the first step. Finally, by categorizing this value into the ranges of risk based on worst possible outcomes provided in Table 7, we are able to provide understandable and actionable insight with respect to the most likely risk escalation scenario.

These steps are illustrated for two simple construction situations in Table 8. For comparison, we also show the range of risk based on real outcomes (provided in Table 5) in which $x_0$ falls.

**Table 8. Illustration of the proposed risk escalation estimation technique**

| attributes | Step 1: PRIOR EVIDENCE<br>risk based on real outcomes ($x_0$) and associated range* | | Step 2: CONDITIONAL QUANTILE ESTIMATE<br>estimate $Q(0.8)$ of risk based on worst potential outcomes and associated range** | |
|---|---|---|---|---|
| hazardous substance, confined workspace | $590 + 115 = 705$ | Extreme | 7266 | Extreme |
| hammer, lumber | $5 + 53 = 58$ | Medium | 676 | High |
| hand size pieces | 7 | Low | 145 | Low |

* based on the ranges proposed in Table 5

** based on the ranges proposed in Table 7

## LIMITATIONS

Since the entire process of computing risk values is data driven, the final risk values of the attributes are expected to change from one injury report database to another, and from one set of exposure values to another, even though the distributions of safety risk based on real and worst potential outcomes are expected to remain the same (i.e., heavy-tailed). Also, in this study, we used a rather small dataset (final size of 814 injury reports) to provide a proof of concept for our methodology. With larger datasets, more attributes would play a role in characterizing the different regimes presented in Figure 4, and their respective signature would therefore enjoy a higher resolution.

## CONCLUSIONS AND RECOMMENDATIONS

In the first part of this paper, we proposed a methodology to compute univariate and bivariate construction safety risk from attributes and outcomes extracted from raw textual injury reports (i.e., candid observations of the jobsite at injury time). We then showed the empirical probability distribution of construction safety risk to be strikingly similar to that of earthquake, ocean waves, asteroid impact, flood magnitude, and other natural phenomena. Motivated by this finding, we posited that construction safety risk may benefit from being studied in a fully empirical fashion, and introduced data-driven, nonparametric univariate and bivariate modeling and stochastic simulation schemes.

Our approaches were inspired by the state-of-the-art in hydroclimatology and insurance, and are respectively based on Kernel Density Estimators and empirical Copulas. Our nonparametric and empirical data-driven techniques are free of any model fitting, parameter tuning, or assumption making. Therefore, they can be used as a way to ground risk-based safety-related decisions under uncertainty on objective empirical data far exceeding the personal history of even the most experienced safety or project manager. Additionally, the combined use of the attribute-based framework and raw injury reports as the foundation of our approach allows the user to escape the limitations of traditional construction safety risk analysis techniques that are segmented and rely on subjective data. Finally, the attribute-based nature of our methodology enables easy integration with building information modeling (BIM) and work packaging.

We believe this study gives promising evidence that transitioning from an opinion-based and qualitative discipline to an objective, empirically grounded quantitative science would be highly beneficial to construction safety research. Just like the accurate modeling and simulation of natural phenomena such as streamflow, precipitation or wind speed is indispensable to successful structure dimensioning or water reservoir management in Civil engineering, the underlying assumption is that improving construction safety calls for the accurate quantitative modeling, simulation, and assessment of safety risk.

One interesting finding obtained on the data set we used to test our methodology is that central risk shapers are attributes related to high energy levels (e.g., hazardous substance, machinery, forklift) and to human behavior (e.g., improper security of tools, lifting/pulling/manual handling). We remind the reader that the risk values based on real and worst potential outcomes are reported for all attributes in Table 1.

The analyst should decide whether to split the injury report database based on industry branches in which the company is involved, and whether to consider overall exposure values or exposure values per discipline. In any case, interpretations of the risk scores remain valid as long as they are made within the domain from which originated the reports and the exposure values. The former allows to identify differences in risk profiles from one industry discipline to another and to obtain a final product tailored to a particular branch. On the other hand, the latter gives the big picture at the overall company level.

Also, there is currently no automated way to extract real and worst possible severity from a given textual injury report, and it is therefore necessary to have human coders perform the task, which is a costly and lengthy process. Future research should address this issue.

**AKNOWLEDGMENTS**

Sincere thanks go to Prof. Arthur Charpentier for his kind help on nonparametric copula density estimation and on the bivariate smoothed bootstrap, and to Prof. Carl Scarrott for the insights provided on dynamic mixture modeling.